\documentclass{article}
\usepackage{spconf,amsmath,graphicx}
\usepackage{amsfonts}
\usepackage{subcaption}
\usepackage{multirow}
\usepackage{booktabs}

\title{An investigation of distribution alignment in multi-genre speaker recognition}
\name{Zhenyu Zhou$^{1,3}$, Junhui Chen$^{3}$, Namin Wang$^2$, Lantian Li$^3$, Dong Wang$^1$}
\address{$^1$Center for Speech and Language Technologies, BNRist, Tsinghua University, China \\
         $^2$Huawei Cloud, China~~~~~$^3$Beijing University of Posts and Telecommunications, China}
\begin{document}
\ninept
\maketitle
\begin{abstract}

Multi-genre speaker recognition is becoming increasingly popular due to its ability to better represent the complexities of real-world applications.
However, a major challenge is the significant shift in the distribution of speaker vectors across different genres.
While distribution alignment is a common approach to address this challenge,
previous studies have mainly focused on aligning a source domain with a target domain, and the performance of multi-genre data is unknown.

This paper presents a comprehensive study of mainstream distribution alignment methods on multi-genre data, where multiple distributions need to be aligned.
We analyze various methods both qualitatively and quantitatively. Our experiments on the CN-Celeb dataset show that within-between distribution alignment (WBDA)
performs relatively better. However, we also found that none of the investigated methods consistently improved performance in all test cases.
This suggests that solely aligning the distributions of speaker vectors may not fully address the challenges posed by multi-genre speaker recognition.
Further investigation is necessary to develop a more comprehensive solution.

\end{abstract}
\begin{keywords}
multi-genre, distribution alignment, speaker recognition
\end{keywords}

\section{Introduction}
\label{sec:intro}

Speaker recognition aims to verify the claimed identity of a speech segment.
After decades of research, contemporary speaker recognition systems have achieved significant advancement,
partly due to the continuous data accumulation and the prevalence of speaker embedding models based on deep neural networks (DNNs).
The x-vector architecture and its variants are among the most widely used speaker embedding models~\cite{snyder2018x,snyder2019speaker,garcia2019x}.
Recently, with carefully designed architectures and training techniques,
deep embedding models have achieved state-of-the-art performance in numerous benchmark evaluation tasks~\cite{sadjadi20222021,huh2023voxsrc}.

Despite the significant progress made, performance on multi-genre data is still poor, especially in the
cross-genre test, i.e., enroll in one genre and test in another.
Analysis showed that this poor performance is largely attributed to the significant shift in the distribution of 
speaker embedding vectors from one genre to another~\cite{li2022cn}.
A large body of research has been conducted to address this problem.
One of the widely adopted approaches is domain alignment (DA), which seeks to
minimize the discrepancy of distributions of speaker vectors in a source domain and a target domain.
Depending on whether speaker labels are available,
existing DA algorithms can be categorized into unsupervised DA (UDA) and supervised DA (SDA).
UDA does not require speaker labels and tries to match the \emph{overall} distributions of two domains.
Correlation alignment (CORAL)~\cite{sun2016return,lee2019coral+,li2022coral++} and
its deep version DeepCORAL~\cite{sun2016deep} are among the most popular methods.
These methods focus on aligning the second-order statistics, i.e., covariance matrices, of the speaker vectors
in the target domain and source domain.
Maximum mean discrepancy (MMD)~\cite{borgwardt2006integrating,lin2020multi,wang2021multi} is another representative method, which
employs the radial basis function (RBF) kernel to measure the distance of two distributions.

For SDA, speaker labels are known so that the between-speaker distribution and within-speaker distribution can be
separately matched. WBDA (within-between distribution alignment) proposed by Hu et al.~\cite{hu2022class} recently
belongs to this category and showed promising results. The key idea of WBDA is to align the within-class \emph{and} between-class distributions of two domains separately, and the matching losses are combined by weighted average.
Additionally, if some speakers appear in both domains, additional and strong information is available for domain alignment. The most straightforward way
to use this information is to enforce the speaker vectors of the same speaker but from different domains close to each other, for instance by imposing a center loss~\cite{wen2016discriminative,li2018deep}.

While DA methods have been extensively studied, previous work
primarily focused on aligning two domains, overlooking the fact that most real applications work in multi-domain conditions. For instance, one may want to unlock his mobile by voice
in his office, on the street, or in his car. Addressing this real-world complexity requires multi-domain alignment.
In addition, most of the existing DA work focuses on the single-domain test, largely due to the lack of cross-domain data, though the cross-domain test is ubiquitous and more crucial in practice, e.g., enroll in a quiet office but test in a noisy cafeteria.

In this paper, we focus on multi-domain alignment, which is supposed to be more challenging than
the previously studied two-domain alignment. 
We use the CN-Celeb dataset~\cite{fan2020cn,li2022cn} to perform the study. The dataset contains 11 genres that cover most of the scenarios a real application may be deployed to.
We treat each genre as a particular domain and design multi-domain DA algorithms that can align all the genres. Hereafter in this paper, `domain' and `genre' will be used interchangeably.

The contribution of the paper is two-fold: (1) We extend existing popular two-domain DA algorithms to tackle multi-domain DA tasks, which is more difficult and has not been explored so far; (2) We present a comparative study on the multi-domain DA methods with a multi-genre speaker verification task, in particular on cross-genre tests that are more concerned in real applications. 
Our experiments show that all the investigated DA methods outperform the baseline, highlighting their effectiveness.
However, none of these DA methods consistently improve performance in all test cases,
indicating that their generalizability is still limited and solely aligning the distributions across different
genres cannot fully address the challenges posed by multi-genre and cross-genre tests.
As far as we know, this work is among the earliest and most systematic studies on multi-domain DA.

\section{Method}
\label{sec:method}

\subsection{Genre sampling}

We first formulate the training scheme. 
Denoted the training objective by $\mathcal{L}$ and it consists of two terms:
one is the speaker classification loss, which is cross-entropy in our case and is denoted by $\mathcal{L}_{\mathrm{CE}}$,
and the other is the regularization term attributed to various DA methods, denoted by $\mathcal{L}_{\mathrm{DA}}$:

\begin{equation}
\mathcal{L} = \mathcal{L}_{\mathrm{CE}} + \lambda \mathcal{L}_{\mathrm{DA}},
\end{equation}
\noindent where $\lambda$ is a hyper-parameter to control the contribution of $\mathcal{L}_{\mathrm{DA}}$.

A key problem that needs to be addressed is that most existing DA algorithms are
designed to align two domains. In order to extend the two-domain DA algorithms to 
tackle multi-domain DA, we designed a genre sampling approach, which samples 
two genres for each mini-batch and aligns them using the conventional DA algorithms.
The hope is that with lots of mini-batches accumulated, the two-domain
alignment on random genre pairs allows aligning all the domains together. 
This genre sampling approach has been used to extend DeepCORAL, MMD, and WBDA. 
Another way to align multiple domains is to utilize single-speaker 
multi-genre data. This type of data provides a strong alignment signal 
and can be used to enforce speaker-by-speaker alignment, e.g., by a center loss.  
Details of these methods are described shortly.

\subsection{DeepCORAL}

DeepCORAL matches two genres, denoted by $i$ and $j$, by minimizing the following loss:

\begin{equation}
\mathcal{L}_{\mathrm{DA}}^{\mathrm{CORAL}} = \frac{1}{4d^2} \left\| \varSigma_i - \varSigma_j \right\|_F^2,
\end{equation}
\noindent
where $\varSigma_i$ and $\varSigma_j$ represent the covariances of the speaker vectors of the two genres,
 $d$ the dimensionality of the speaker embedding, and $\left\| \cdot \right\|_F^2$  is the Frobenius norm.

To address the multi-domain alignment problem, two genres are first sampled in a minibatch,
and speakers and utterances are then sampled from the two genres. The distributions of the two
genres are then aligned by the pair-wised DeepCORAL. 

\subsection{MMD}

As in DeepCORAL, two genres are randomly sampled for each mini-batch.
Given two sets of speaker vectors of the two genres, denoted by \( {X}=\{x_i\}^N_{i=1} \) and \( {Y}=\{y_j\}^N_{j=1} \) respectively,
MMD measures the similarity of the two genres by computing the mean squared difference as follows~\cite{gretton2006kernel}:

\begin{small}
\begin{equation}
\begin{split}
D({X},{Y}) &= \frac{1}{N^2} \sum_{i=1}^{N} \sum_{i'=1}^{N} {k(x^i,x^{i'})} -\frac{2}{N^2} \sum_{i=1}^{N} \sum_{j=1}^{N} {k(x^i,y^{j})} \\
&\quad + \frac{1}{N^2} \sum_{j=1}^{N} \sum_{j'=1}^{N} {k(y^j,y^{j'})},
\end{split}
\end{equation}
\end{small}

\noindent where $k(\cdot)$ is a kernel function. In our experiments, the following RBF kernel is used:

\begin{equation}
k({x},{y})=\mathrm{exp}(-\frac{1}{2{\sigma}^2}\left\|{x}-{y} \right\|^2),
\end{equation}
\noindent where \(\sigma\) is the width parameter.

Therefore, the MMD loss for a mini-batch that samples genre $i$ and $j$ is computed as follows:

\begin{equation}
\mathcal{L}_{\mathrm{DA}}^{\mathrm{MMD}} = {D}(X_i, X_j),
\end{equation}

\noindent where $X_i$ and $X_j$ represent the utterances sampled from genre $i$ and genre $j$, respectively.

\subsection{WBDA}

As in DeepCORAL and MMD, two genres are sampled in each mini-batch, and
$S$ speakers are sampled for each genre, with $M$ utterances per speaker.
We then compute the within-class covariance $\varSigma^W_{cov}$ and between-class covariance $\varSigma_{cov}^B$ for each genre:

\begin{equation}
\begin{split}
\varSigma_{cov} &= \varSigma_{cov}^{W} + \varSigma_{cov}^{B} \\
\varSigma_{cov}^{W} &= \frac{1}{SM}\sum_{s=1}^{S} \sum_{i=1}^{M}(x_s^i - \mu_s)(x_s^i - \mu_s)^T \\
\varSigma_{cov}^{B} &= \frac{1}{SM}\sum_{s=1}^{S}n_s(\mu_s - \mu)(\mu_s - \mu)^T
\end{split}
\end{equation}

\noindent where for speaker $s$, $x_s^i$ is the $i^\text{th}$ utterances, $n_s$ is the number of utterances, and $\mu_s$ is the speaker center.
$\mu$ is the global center of all speakers. It was found that normalizing the covariance matrices to correlation matrices leads to better performance. Following~\cite{hu2022class}, the covariances are normalized as follows:


\begin{equation}
\begin{split}
\varSigma^W &= {\varSigma_{cov}^{W}}/{\sqrt{\mathrm{Diag}(\varSigma_{cov}^{W}) \mathrm{Diag}(\varSigma_{cov}^{W})^T}}\\
\varSigma^B &= {\varSigma_{cov}^{B}}/{\sqrt{\mathrm{Diag}(\varSigma_{cov}^{B}) \mathrm{Diag}(\varSigma_{cov}^{B})^T}}
\end{split}
\end{equation}

Finally, the WBDA loss for a mini-batch that samples genre $i$ and $j$ is computed as follows:

\begin{equation}
\mathcal{L}_{\mathrm{DA}}^{\mathrm{WBDA}} = \alpha \lVert \varSigma_{i}^W - \varSigma_{j}^W \rVert_F^2 + \beta \lVert \varSigma_{i}^B - \varSigma_{j}^B \rVert_F^2,
\end{equation}
\noindent where \(\rVert \cdot \rVert_F^2\) denotes the squared matrix Frobenius norm.
$\alpha$ and $\beta$ are used to balance the weights of the two distributions.

Compared to DeepCORAL and MMD, WBDA factorizes the global distribution into within-class distribution $\varSigma^W$
and between-class distribution $\varSigma^B$.
On one hand, WBDA makes full use of the value of the speaker label of each genre so that it can obtain more precise distribution statistics.
On the other hand, it is more flexible to balance the contributions of within-class and between-class distributions.

\subsection{Center loss}

The center loss aims to reduce the distances between the speaker vectors of the same speaker but in different genres. It utilizes single-speaker multi-genre data, rather than genre sampling. Specifically, we sample $S$ speakers in each mini-batch, with
$M$ utterances per speaker. 
Note that the utterances are sampled without knowing the genre information. 
However, most of the speakers in CN-Celeb appear in multiple genres, so the data 
in the mini-batch naturally provides strong signals for multi-domain alignment. 
The center loss is implemented as follows:

\begin{equation}
\mathcal{L}_{\mathrm{DA}}^{\mathrm{CenterLoss}} = \frac{1}{SM} \sum_{s=1}^{S} \sum_{i=1}^{M} \|x_s^i - \mu_{s}\|^2,
\end{equation}

\noindent where $x_s^i$ denotes the $i^{\text{th}}$ utterances from speaker $s$ and $\mu_s$ represents the center vector of speaker $s$, computed
by averaging all the speaker vectors belonging to speaker $s$ in the mini-batch.

\section{Experiments}
\label{sec:experiments}

In this section, we test the multi-domain DA methods described 
in the previous section on a multi-genre speaker verification task.
Note that in addition to the basic baseline system that involves no domain alignment, 
we also implement a popular domain-invariant learning method,
i.e., multi-domain adversarial training (DAT)~\cite{ganin2015unsupervised,wang2018unsupervised,wang2021adversarial} as a more competitive baseline.

\subsection{Data and Settings}

\emph{CN-Celeb}~\cite{li2022cn}, a large-scale multi-genre speaker recognition dataset, was used in our experiments.
It contains more than 600k utterances from 3,000 Chinese celebrities, and the utterances cover 11 different genres in the real world.
More importantly, it consists of a substantial amount of single-speaker multi-genre data, making it suitable for comparing
the center loss approach and other DA methods.
In our experiments, we employed the CN-Celeb1.dev and CN-Celeb2 for model training, and the CN-Celeb1.eval for evaluation.
The data profile of the training and evaluation followed the evaluation protocol of CNSRC 2022\footnote{http://cnceleb.org/competition}.

We employed the baseline system of CNSRC 2022\footnote{https://gitlab.com/csltstu/sunine} to construct the speaker embedding model.
It adopts the ResNet34 structure~\cite{he2016deep} with squeeze-and-excitation (SE) layers~\cite{hu2018squeeze} and
attentive statistics pooling (ASP)~\cite{okabe2018attentive}. The model was trained
using the additive angular margin Softmax (AAM-Softmax) loss~\cite{xiang2019margin}.
Once trained, the 256-dimensional activations of the last fully connected layer are read out as an x-vector.
The simple cosine distance is used to score the trials.
Besides, in order to achieve more state-of-the-art (SOTA) performance, we also employed some efficient and effective
training-optimization strategies, such as data augmentation~\cite{snyder2018x}, short-utterance combination~\cite{chen2021self}, Adam optimizer with learning rate warm-up,
margin training scheduler~\cite{wang2022wespeaker}, etc.
The source code can be found at\footnote{https://github.com/buptzzy2018/DA\_MG}.

\subsection{Basic results}
\label{sec:basic}

We first present the basic results evaluated on CN-Celeb1.eval with various DA methods.
The results in terms of equal error rate (EER) are reported in Table~\ref{tab:basic}, including 
the baseline, DAT, DeepCORAL, MMD, CenterLoss, and WBDA. Note that WBDA involves aligning 
both the within-speaker distribution and between-speaker distribution. To view the contribution of these 
two alignments, we also present the results of within-speaker distribution alignment (WDA) and between-speaker distribution alignment (BDA).



\begin{table}[htb!]
  \caption{EER(\%) results on CN-Celeb1.eval with various DA methods. $\lambda$ is the hyper-parameter in Eq.(1).}
  \label{tab:basic}
  \centering
  \begin{tabular}{l|l|c|c}
	\toprule[1pt]
		Type & Method & $\lambda$ & EER(\%)  \\
	\midrule[1pt]
		\multirow{2}{*}{Baseline}  & ResNet34 & - & 8.724 \\
                                       & \quad + DAT & 0.0006 & 8.702 \\
	\midrule[1pt]
		\multirow{2}{*}{UDA} & \quad + DeepCORAL & 0.1 & 8.538 \\
                                 & \quad + MMD & 0.8 & 8.561 \\
	\midrule[1pt]
		\multirow{4}{*}{SDA} & \quad + CenterLoss & 0.1 & 8.437 \\
								   & \quad + WDA & 0.9 & 8.369 \\
									& \quad + BDA & 0.03 & 8.568 \\
									& \quad + WBDA & 0.9 & \textbf{8.308} \\
	\bottomrule[1pt]
  \end{tabular}
\end{table}

\begin{figure}[h]
 \centering
 \includegraphics[width=\linewidth]{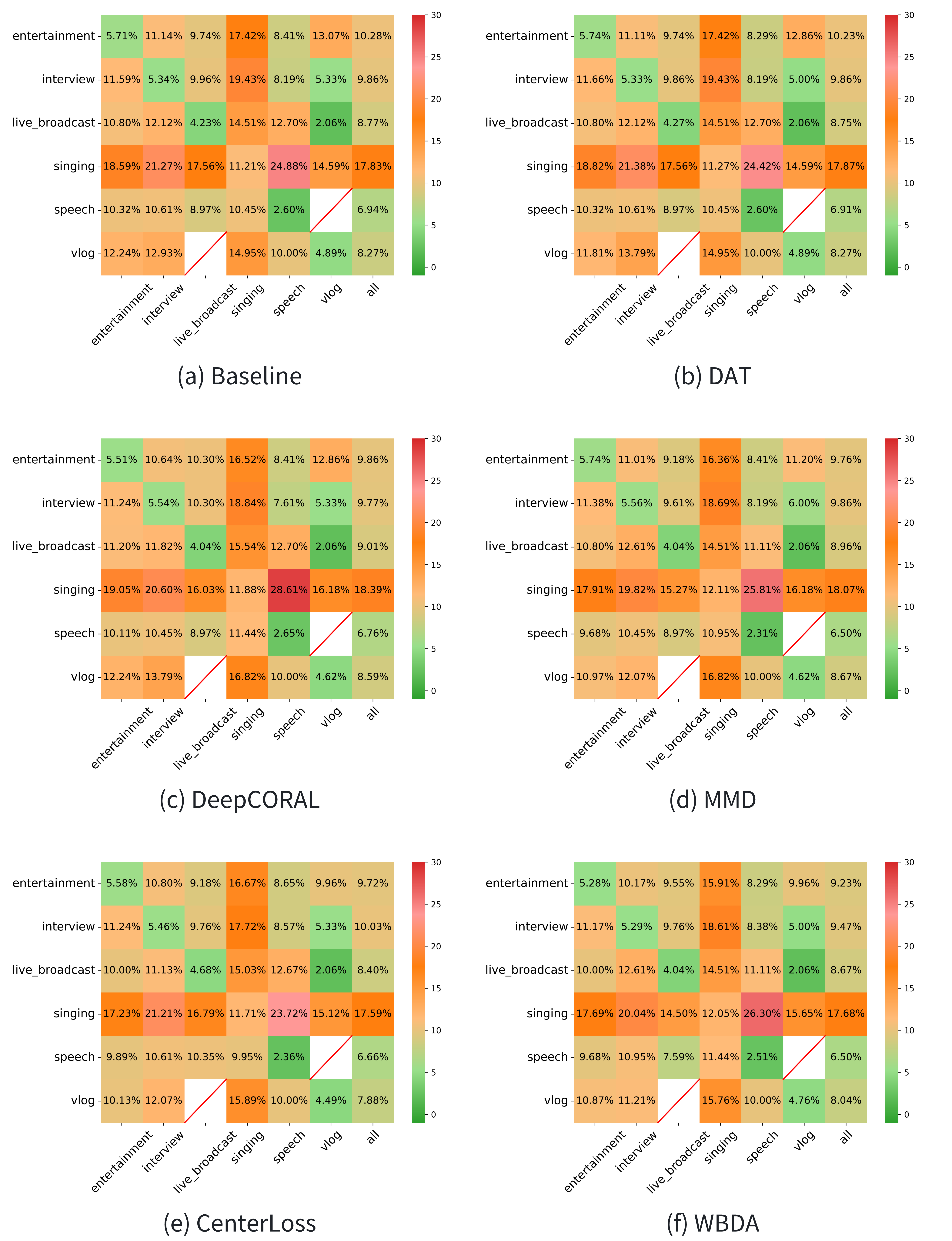}
 \caption{Cross-genre tests with different methods.}
 \label{fig:cross}
\end{figure}

Firstly, it can be observed that all these DA methods offer better performance than the baseline system
and the more competitive DAT method.
This demonstrates that the DA methods that were originally designed for two-domain alignment can be 
also used to align multiple domains using the genre sampling approach, and the alignment leads to 
notable performance improvement in multi-genre verification tasks. 

Secondly, SDA methods consistently outperform UDA methods. This is expected as speaker labels are used in 
SDA but ignored in UDA: for CenterLoss, they are used to align the same speaker in different genres; for 
WBDA, they are used to decompose with-speaker and between-speaker distributions and align them separately.  
Comparing WDA and BDA, one can find that WDA is more effective and contributes more to the success of WBDA.
An unexpected observation is that CenterLoss performs worse than WBDA, although it uses single-speaker 
multi-genre data. Summarizing these results, it seems that enforcing a homogeneous shape for 
the within-speaker distributions in different genres 
(Note correlation matrix defines the shape of a distribution partly) is mostly effective in aligning different genres.
This within-speaker DA is even more effective than pushing
speaker vectors in different genres together, although the latter uses
a stronger alignment signal. Nonetheless, this does not mean 
that single-speaker multi-genre data is useless; 
it just suggests that more appropriate methods should be investigated.



\subsection{Cross-genre tests}

To provide a more comprehensive comparison of these DA methods,
we focus on cross-genre tests and compute genre-to-genre performance matrices.
Firstly 6 largest genres in terms of the number of speakers and utterances are selected
to construct the cross-genre test trials.
The enrollment utterance involves 20 seconds of speech, and the rest utterances are used for testing.

Figure~\ref{fig:cross} shows the results. 
The plots in this figure represent the EER results with different methods.
For each plot, the number and the color at each location represent the EER results 
under the enrollment genre corresponding to its row and the test genre corresponding to its column,
and the last column shows the overall results that the enrollment is based on one genre and the test is on all the genres.

Firstly, it can be found that all the DA methods surpass the baseline (a) and DAT (b) in most cross-genre cases,
further indicating the effectiveness of these DA methods.
When comparing different DA methods, WBDA shows some advantages.
Specifically, in the 6 cross-genre tests, WBDA offers a relative EER reduction larger than 10\%,
including entertainment-vlog (23.79\%), singing-live\_broadcast (17.43\%), and speech-live\_broadcast (15.38\%).

Secondly, it can be seen that different methods exhibit their own advantages in different test scenarios,
For instance, CenterLoss wins at the interview-singing test, MMD wins at the interview-live\_broadcast test,
DeepCORAL wins at the interview-speech test, and WBDA wins at the interview-vlog test.

In summary, none of the present DA methods can consistently outperform the baseline.
This suggests that solely aligning the distributions of multi-genre speakers might not fundamentally
address the complex variations in multi-genre and cross-genre scenarios.

\subsection{Visualization}

\begin{figure}[t]
  \centering
  \includegraphics[width=0.95\linewidth]{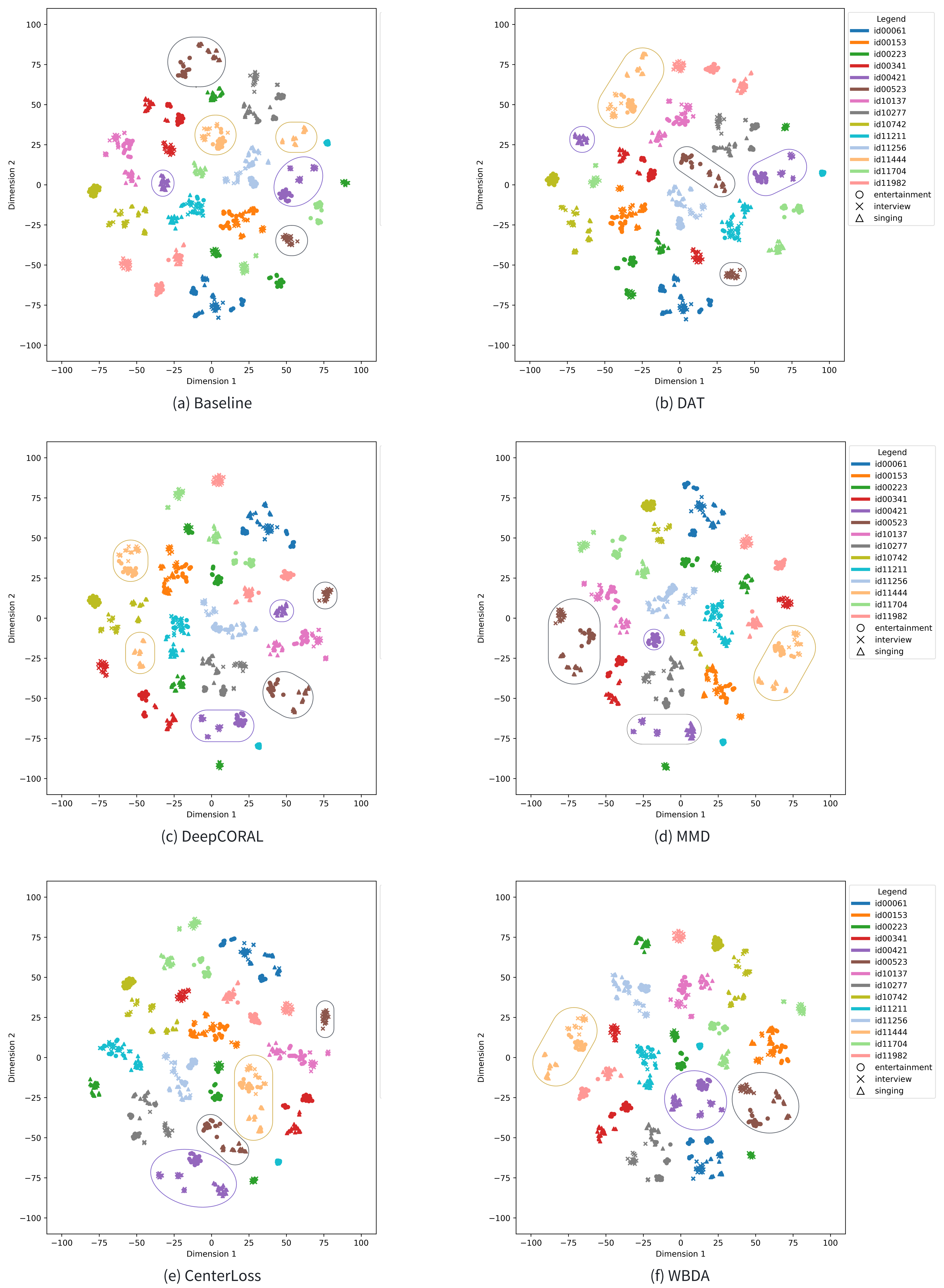}
  \caption{Multi-genre speaker distributions with different methods plotted by t-SNE.}
  \label{fig:tsne}
\end{figure}

In this section, we present a visualization study to demonstrate the 
contribution of different DA methods. 
We selected utterances of 14 speakers from 3 genres to perform the visualization. 
With each model, the speaker vectors of these utterances are first produced, and then 
the t-SNE toolkit~\cite{saaten2008} is applied to project the speaker vectors to a 2-dimensional space
so that can be viewed as a picture. 
Fig.~\ref{fig:tsne} presents the results, where each plot shows the distribution of the speaker vectors generated by 
a particular model, and in each plot, each color represents a speaker and each shape represents a genre.

For the baseline system, we can observe that the within-class distributions for most speakers are highly dispersed,
showing clear genre dependency.
In other words, vectors from the same speaker in three genres tend to cluster into three groups,
as demonstrated by the three speakers in the circles. 

With the DA methods, this genre dependency is reduced, as demonstrated by the more compact clusters being circled out.
This is more clear in the WBDA plot, where some broadly dispersed speaker vectors are aggregated together. 
This demonstrates that through distribution alignment, the model can learn more genre-invariant speaker representations.
Nevertheless, the intra-speaker variation caused by genre change is still noticeable, indicating that distribution shift 
still remains a challenging problem for multi-genre speaker recognition, and requires deep and lasting investigation. 

\section{Conclusion}

This paper extends existing two-domain distribution alignment (DA) methods to tackle multi-domain distribution alignment tasks. In particular, we focus on multi-genre speaker recognition, where each genre is a specific domain, 
and the enrollment and test data may be from one of several genres, hence suffering from severe distribution shifts. 
We propose a genre sampling approach, which randomly chooses two genres in each mini-batch, 
and performs two-domain distribution alignment. This genre sampling approach has been demonstrated to 
be an effective way to perform multi-domain distribution alignment. 

Our experiments were conducted with CN-Celeb which involves 11 genres of speech data. 
Several state-of-the-art DA methods were comparatively studied, 
and the results demonstrated that all the investigated DA methods can be well
extended to align multiple domains, using the genre sampling approach. 
They yield clear performance improvement compared to the baseline in our experiments,
especially the WBDA approach that aligns both the within-speaker and between-speaker 
distributions. 
However, none of the investigated methods consistently improve performance across all test cases, and different methods showcase their respective advantages in different test cases.
These observations indicate that multi-genre speaker verification remains difficult 
and deserves dedicated study. 


\newpage
\bibliographystyle{IEEEbib}
\bibliography{refs}

\end{document}